# Direct Visualization of Electric Field Induced Structural Dynamics in Monolayer Transition Metal Dichalcogenides

*Akshay A. Murthy†,§, Teodor K. Stanev‡, Roberto dos Reis†,◊, Shiqiang Hao†, Christopher Wolverton†, Nathaniel P. Stern‡, Vinayak P. Dravid\*,†,§,◊*

†Department of Materials Science and Engineering, §International Institute for Nanotechnology (IIN), ‡Department of Physics and Astronomy, and ◊Northwestern University Atomic and Nanoscale Characterization Experimental (NU*ANCE*) Center, Northwestern University, Evanston, Illinois 60208, USA

\*Corresponding author

Vinayak P. Dravid:  v-dravid@northwestern.edu

**Abstract** : Layered transition metal dichalcogenides (TMDs) offer many attractive features for next-generation low-dimensional device geometries. Due to the practical and fabrication challenges related to *in situ* methods, the atomistic dynamics that give rise to realizable macroscopic device properties are often unclear. In this study, *in situ* transmission electron microscopy techniques are utilized in order to understand the structural dynamics at play, especially at interfaces and defects, in the prototypical film of monolayer $MoS_2$ under electrical bias. Through our sample fabrication process, we clearly identify the presence of mass transport in the presence of a lateral electric field. In particular, we observe that the voids present at grain boundaries combine to induce structural deformation. The electric field mediates a net vacancy flux from the grain boundary interior to the exposed surface edge sites that leaves molybdenum clusters in its wake. Following the initial biasing cycles, however, the mass flow is largely diminished, and the resultant structure remains stable over repeated biasing. We believe insights from this work can help explain observations of non-uniform heating and preferential oxidation at grain boundary sites in these materials.

In recent years, defect engineering of layered transition metal dichalcogenides (TMD) has served as a popular paradigm for constructing unique devices with enhanced functionalities. In particular, point and line defects serve as an effective pathway for tuning the electronic,[1,2] optical,[3,4] and catalytic properties[5,6] in these materials. For example, grain boundaries (GBs) in TMDs formed through vapor phase synthesis demonstrate electronic properties that differ from the surrounding regions,[7,8] as well as, electronic structures that can be modulated by systematically varying the GB tilt angle.[9,10] These interfaces have been used in devices that demonstrate electroluminescent behavior,[11] as well as gate tunable memristive behavior, making them candidate systems for neuromorphic computation.[12,13] Recent reports, however, have indicated that TMD grain boundaries serve as sites for preferential heating and oxidation during device operation.[14-16] To date, these phenomena have been generally attributed to the presence of nanoporous regions[14] and sulfur vacancies,[15,16] but dynamic high-resolution techniques are essential in order to fully correlate these structures to observable properties.

To better explore these relationships, one particularly useful methodology is *in situ* electrical biasing TEM. Through this approach, the electrical response of the material can be effectively monitored and correlated directly to nanophysical phenomena. As such, this methodology has achieved significant popularity through electrochemical studies aiming to understand structural dynamics at the electrode interfaces,[17-19] as well as investigations focused on acquiring a mechanistic understanding of electronically-induced resistive switching.[20-22] Nonetheless, due to the practical challenges linked to directly probing atomically thin materials using traditional *in situ* biasing sample methods, there have only been a couple of reports investigating monolayer TMDs using this approach.[23,24]

In this study, we develop a fabrication methodology that combines electrical biasing with transmission electron microscopy. This approach allows us to investigate the structural dynamics at play when polycrystalline films of monolayer $MoS_2$ are electrically biased. We find that when an electric field is applied, strained regions around the grain boundary act as vacancy sources and voids at the grain boundary act as vacancy sinks, leading to a net vacancy flux towards the grain boundaries. Additionally, we find that this vacancy flow process appears to yield regions of molybdenum clusters that aggregate near the voids. Finally, we find that the vacancy flow process does not continue indefinitely, but rather after a few initial biasing cycles, the material reaches a stable state. Assessment of polycrystalline $MoS_2$ using *in situ* electrical biasing TEM gives the ability to draw direct insights that help us explain the origin of preferential localized heating and oxidation at grain boundary regions that have been demonstrated previously.

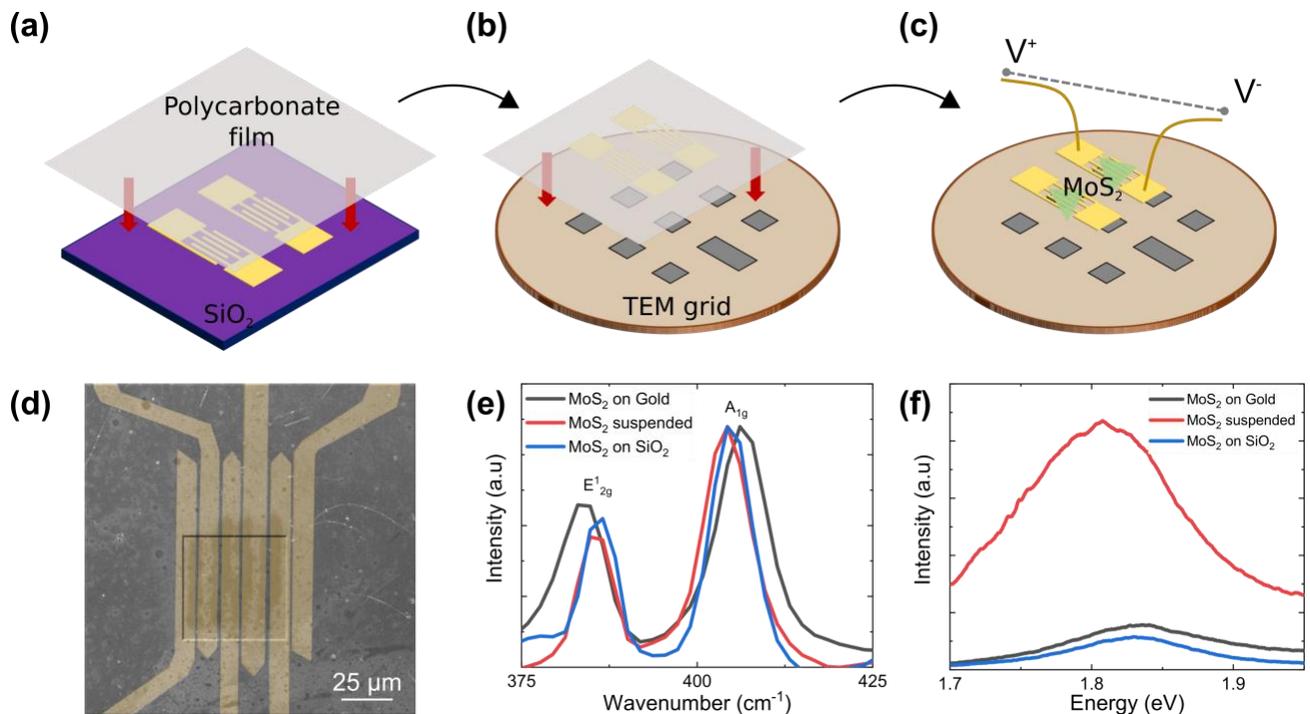

**Figure 1. TEM device fabrication**. **(a-c)** Schematic of fabrication process. Metal electrodes are first picked up from an arbitrary substrate using a polymer support film **(a)**. These electrodes are then placed on a TEM grid **(b)**. Polycrystalline monolayer $MoS_2$ film is then picked up and placed in an identical fashion **(c)**. **(d)** Colorized SEM image of fabricated device. The gold color is indicative of the metal electrodes and the beveled region is indicative of the region where $MoS_2$ is suspended. Since polycrystalline $MoS_2$ is used, multiple grain boundaries are present in the regions between the electrodes. **(e)** Raman spectra taken from $MoS_2$ regions supported on $SiO_2$, supported on gold, and that are suspended. Due to differences in charge doping and strain between these regions, the positions of the Raman modes vary with substrate. (f) Photoluminescence (PL) spectra taken from the same $MoS_2$ regions. Due to the inherent differences in charge doping and strain, the position and magnitude of the PL peaks vary with substrate.

**Results/Discussion**

In order to prepare a TEM sample for *in situ* electrical biasing, 75 nm gold interdigitated electrodes with a 1 μm spacing are defined on a 300 nm $SiO_2$/Si substrate. Using a polycarbonate transfer technique described previously,[25] the electrodes were picked up and placed onto a $SiO_2$/Si TEM grid such that they covered a pre-defined aperture and created an interdigitated material support bridge (Figure 1a). Subsequently, a polycrystalline film of monolayer $MoS_2$, synthesized through vapor phase techniques as described in previous reports,[14, 26] is then picked up and stacked on top of the gold electrodes using the same polycarbonate stamp technique (Figure 1b). This method allows for monolayer $MoS_2$ to be suspended between adjacent pairs of electrodes. Finally, wires are bonded between the contact pads and the TEM mount contact points on a Nanofactory holder (Figure 1c). This approach thus allows for applying lateral electric fields across each pair of interdigitated electrodes. A scanning electron microscopy (SEM) image of the interdigitated electrodes and $MoS_2$ stack is seen in Figure 1d.

A comparison of the Raman and photoluminescence (PL) spectra taken from three distinct regions is provided in Figures 1e and 1f, respectively. These include spectra taken from the region where $MoS_2$ is supported by $SiO_2$ around the aperture, a region where $MoS_2$ is supported by the gold electrodes, and a region where $MoS_2$ is suspended. Relative to the suspended region ($E_{2g}^1$ =384.8 $cm^{-1}$), the $E_{2g}^1$ mode is red-shifted when $MoS_2$ is supported by $SiO_2$ ($E_{2g}^1$ =386.6 $cm^{-1}$) and blue-shifted when $MoS_2$ is supported by the gold electrodes ($E_{2g}^1$ =383.0 $cm^{-1}$). With respect to the out-of-plane Raman mode, relative to the suspended region ($A_{1g}$ =404.3 $cm^{-1}$), the $A_{1g}$ mode is red-shifted when $MoS_2$ is supported by the gold electrodes ($A_{1g}$ =406.1 $cm^{-1}$) and no discernible change is observed when the $MoS_2$ supported by $SiO_2$ ($A_{1g}$ =404.3 $cm^{-1}$). Based on the framework we developed previously to correlate the position of Raman modes to biaxial strain and charge doping in $MoS_2$,[26] this observation suggests that the $MoS_2$ supported by the gold electrodes displays the greatest degree of biaxial strain followed by the suspended MoS$_2$ and then the MoS$_2$ supported by SiO$_2$ region. The biaxial strain measured in the MoS$_2$ layer supported by gold can likely be attributed to the strong interactions between gold and sulfur atoms.[27] In particular, because gold and MoS$_2$ have a large lattice mismatch,[28] these strong interfacial interactions likely induce strain in the monolayer MoS$_2$. Additionally, the moderate level of biaxial strain calculated in the suspended region can be attributed to inevitable sinking of the free-standing monolayer in the unsupported region,[29,30] while the minimal biaxial strain in the MoS$_2$ supported by the SiO$_2$ can be explained by the minimal interlayer interactions observed between transferred monolayer TMDs and amorphous substrates.[26] Similarly, the previously developed framework also indicates that MoS$_2$ supported by the gold electrodes displays the greatest degree of charge doping followed by MoS$_2$ supported by SiO$_2$ region and then suspended MoS$_2$. This is also to be expected as gold and SiO$_2$ substrates have been previously demonstrated to readily dope MoS$_2$.[26, 31] Finally, as shown previously, suspended regions of MoS$_2$ exhibit significantly greater photoluminescence intensities compared to supported regions on SiO$_2$ or gold.[32] The broad emission from the suspended material can be explained by a combination of moderate strain levels and reduced charge doping in this region.[26]

Low magnification and high-resolution transmission electron microscopy images of the materials prior to electrical biasing are seen in Figures 2a-2c. A Fast-Fourier transform (FFT) from Figure 2c is provided as inset and depicts the six-fold symmetry and the expected lattice spacings for MoS$_2$. These images were taken using a Gatan K3-IS direct electron detector, which allows for imaging of the lattice using a reduced electron total dose on the order of 350 e$_-$/Å$_2$. Though a slight degree of hydrocarbon accumulation is seen from the diffuse ring in the FFT pattern, this low dose allows for *in situ* imaging of sensitive materials without inducing significant build-up of hydrocarbon residue and/or introducing structural artifacts.

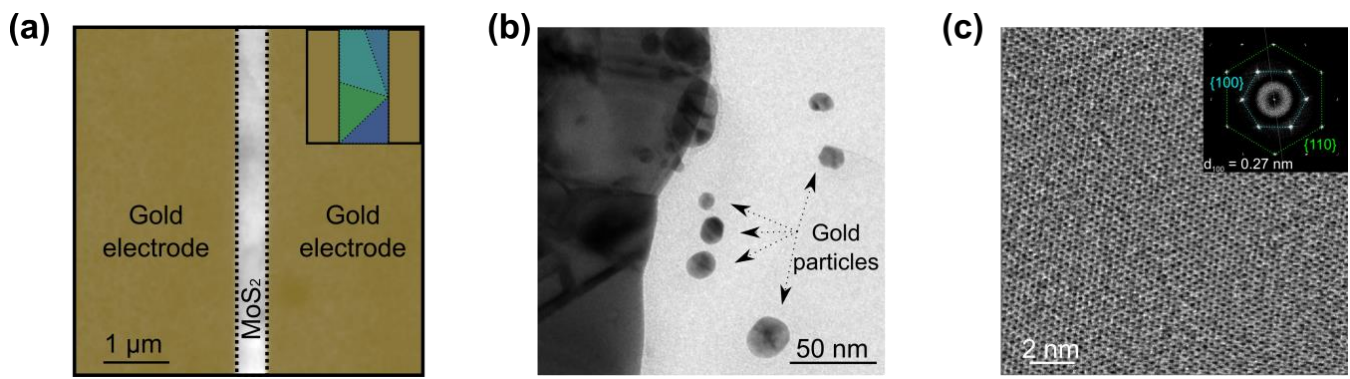

**Figure 2. TEM analysis of MoS$_2$. (a)** Low magnification TEM image of suspended material. Gold region corresponds to metal electrodes and area between dotted lines represents MoS$_2$. Inset provides a schematic representation of the region. In this inset, gold once again corresponds to the electrodes, while the different hues of blue are indicative of MoS$_2$ grains with different orientations and multiple grain boundaries are present. **(b)** Low magnification TEM image of the electrode (left)/ MoS$_2$ (right) interface. Gold nanoparticles are found to be embedded in the MoS$_2$ matrix. **(c)** High resolution TEM image of MoS$_2$ region with a corresponding FFT provided as an inset.

From the representative low magnification TEM images (Figure 2b), we also observe significant migration of nanoparticles up to 100 nm from the contact-material interface. The chemical and structural nature of these gold particles is confirmed through a selected area electron diffraction (SAED) pattern provided in Figure S2. Density functional theory calculations (Figure S3) suggest that this observation is a result of a low migration energy barrier (0.1 eV) for gold atoms on MoS$_2$, which is comparable with the diffusion energy barrier predicted for Li ions on MoS$_2$.[33] This favorable diffusion process is mediated through hopping along adatom sites and likely originates from annealing of the MoS$_2$-gold electrode TEM structure at 200°C prior to electrical testing. The observation of this aspect in every device examined in this study is significant for field-effect transistor MoS$_2$ geometries as it highlights the atomistic dynamics occurring at the metal electrode/MoS$_2$ interface that need to be considered during electronic measurements. Namely, the potential for migration of gold atoms across MoS$_2$ suggests a complicated environment at the near-electrode interface where gold nanoparticles can act as vehicles for plasmonic enhancement,[34,35] charge depletion regions,[36] and localized strain in monolayer MoS$_2$[37] As such, the perturbations that these gold nanoparticles induce on the local electronic structure[31] may lead to interfacial coulomb scattering of charge carriers[38]. This aspect would play a major role in short channel, ballistic-length measurements and can potentially help explain reduced mobilities and conductivities in the sub-200 nm channel length regime.[39]

Additionally, a high-resolution image of a grain boundary is seen in Figures 3a-b. The boundary is indicated by the dotted line and a tilt angle of 4 degrees is apparent through the accompanying FFT. At this grain boundary region, we see the presence of multiple voids. This feature has been observed previously[14, 40,41] and has been attributed to the buildup of stress near a grain boundary. This aspect is supported by geometric phase analysis (GPA)[42] of this grain boundary region, which indicates that, indeed, strain is primarily concentrated at the grain boundary interface in the vicinity of the voids (Figure 3c and Figure S5). As such our results agree with the previous strain analysis conducted by Elibol et. al who used the Peierls-Nabarro and the Foreman dislocation models[40] to indicate that void formation at grain boundaries results from stress release in regions with a large density of dislocation cores.

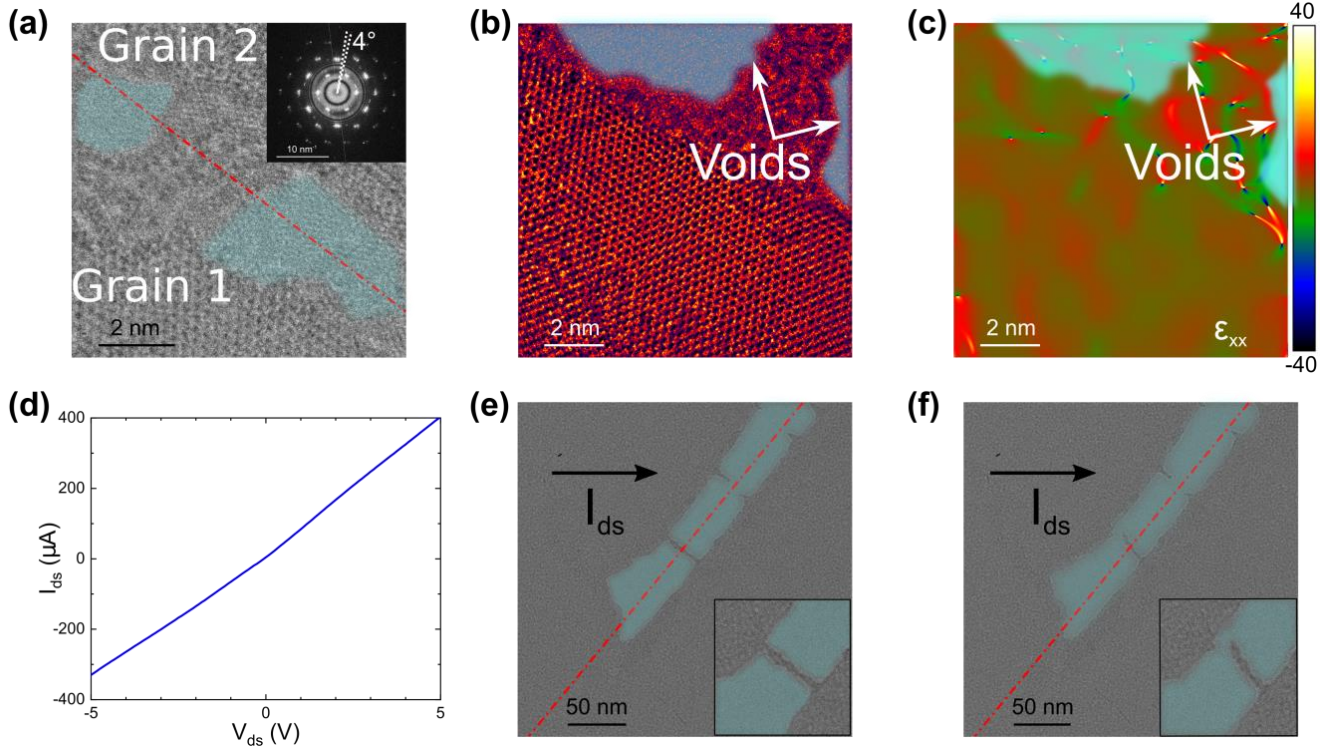

**Figure 3. *In situ* biasing across MoS₂ grain boundary.** Grain boundaries are indicated with red dotted line and voids are colored in blue in all TEM images to aid the eye. **(a)** High resolution TEM image of MoS₂ grain boundary with FFT pattern is provided as an inset. A misorientation angle of 4° is measured. FFT patterns for the individual grains are provided in Figure S4 **(b)** A higher magnification of a void/nanoribbon interface is seen. **(c)** Normal strain in x-direction ($\varepsilon_{xx}$) map of grain boundary region in (b). Strain is highly concentrated near the voids. Additional strain maps for this region are provided in Figure S5. **(d)** Current-voltage response for TEM device. Linearity is indicative of quality contacts. **(e)** Narrow ribbons of material holding together two grains separated by voids prior to biasing. Narrow ribbon of interest is clearly indicated within inset **(f)** Same region following biasing. The narrow ribbon holding together adjacent grains appears to rupture when 1V is applied across the electrodes, as apparent through the inset. Direction of current ($I_{ds}$) is indicated.

Although TMD grain boundaries have been explored heavily for functional electronics and optoelectronics,[11-13] the role these voids play during electrical biasing has not been investigated in detail, despite the fact that the presence of these voids reduces the cross-sectional area for electron conduction. As such, the narrow ribbons of material between adjacent voids are expected to exhibit larger current densities and, in turn, appreciable temperature rises.[14] To explore this aspect, an electrical bias of 1V is applied across adjacent electrodes and the stability of a narrow ribbon with neighboring voids is examined. This electrical potential corresponds to an electric field strength of $1 \times 10^3$ V/cm and is well within the voltage range used for standard device operation.[14] The I-V response for this sample is provided in Figure 3d. A linear response is indicative of a small contact resistance, which has been previously shown to be the case when TMDs are transferred onto existing metal electrodes due to the formation of van der Waal interface largely free of chemical disorder.[43] As is apparent in Figures 3e-f, the application of this electrical potential causes the ribbon to rupture, which suggests that large current densities through these ribbons can be destructive and undesirable (see captured video of this process in SI Video 1). This type of mechanical failure is observed in various regions across the sample including Figure S6a and was only observed when a bias was applied across the sample.

In order to better understand the role pre-existing voids play during electrical transport, low magnification images taken from the grain boundary region are seen in Figures 4a-4b. An image taken from the region prior to applying an electrical potential is seen in Figure 4a, where the voids are indicated in blue. Following an applied potential of 5V across the electrodes, the low magnification image in Figure 4b demonstrates the significant structural change that this region has undergone. Namely, neighboring nanoscale voids appear to have coalesced to form a large crack on the order of a micron. This behavior is seen in multiple regions across various samples as is apparent from Figure S6.

In order to better understand the atomistic dynamics at play when an electrical bias is applied across a grain boundary requires examining the migration energy barriers that molybdenum and sulfur atoms face. Understanding the energetics for this type of process is essential because void growth requires migration of both sulfur and molybdenum atoms from the edges of the void to the interior of the grains. As previous DFT calculations indicate,[44,45] intrinsic line defects present in CVD-grown $MoS_2$ offer a low energy pathway for migration of molybdenum and sulfur atoms. The energy barrier for diffusion through this mode for molybdenum and sulfur atoms is roughly 1.6 eV and between 0.6 and 0.7 eV, respectively. As such, both of these low energy processes are quite accessible when an electrical field is applied across the sample. In addition, based on these calculated energy barriers, it is expected that the migration of sulfur atoms from the edges of the void to the interior of the grain *via* line defects would occur more rapidly than a similar process for molybdenum atoms and thus leave behind a molybdenum-rich region near the void region.

To this end, the representative low magnification TEM image in Figure 4c demonstrates the presence of numerous clusters residing near the void region. EDS confirms the presence of molybdenum in these clusters, which suggests that they are similar to the clusters previously seen by Chen *et al*.[45] during their *in situ* heating experiment (Figure S7). Atomic resolution imaging in that study indicated that these clusters were mainly composed of metal molybdenum nanoparticles, which supports the DFT calculations discussed above. Based on the thermal effects associated with electrical biasing, which will be discussed below, we expect to see similar clusters to those observed by Chen *et al*. when $MoS_2$ grains are heated.[45] As such, we hypothesize that electrical biasing leads to thermally-induced molybdenum and sulfur migration mediated by line defects, which generates long void channels and molybdenum clusters.

Since the Raman modes in MoS$_2$ red-shift due to heating and softening of phonon modes,[46-48] *ex situ* Raman spectroscopy can be used to measure this temperature rise in the suspended material when an electrical bias is applied. In particular, shifts in the position of the A$_{1g}$ mode in MoS$_2$ are linearly related to the average temperature in the material (Figure S8). In order to estimate the temperature rise present in these suspended devices, Raman peak shifts were first calibrated as a function of temperature below room in a cryostat, which showed trends similar to previously reported values in the literature.[49] Using this relationship, the Raman shifts measured in the A$_{1g}$ mode predicted an upper limit in temperature rise of 180 K in these suspended devices (Figure S8). Since even these upper limit values fall within the range of temperature rises previously reported in MoS$_2$ devices supported by a SiO$_2$/Si substrate using optothermal Raman techniques,[49] we believe the observations and insights drawn in these suspended devices can be directly translated to MoS$_2$ devices supported on substrates as well.

Having suggested that the migration process responsible for void growth is driven by thermally induced motion of sulfur and molybdenum atoms *via* line defects, it is worth exploring how these systems behave after a prolonged period. In Figure 4d, the same region shown in Figures 4a-4b is seen after cycled multiple times from 0 to 5V at a rate of 100 mV/s. Additionally, a high-resolution image of a nanoribbon fabricated through an identical cycling process is seen in Figure 4e. Following an electrical bias of 10V, the same region is seen in Figure 4f and displays no discernible structural changes. A video of this process is provided in the SI Video 2. Based on the nearly identical nature of the areas before and after biasing, it is apparent that in this case, the rate of structural changes diminishes precipitously.

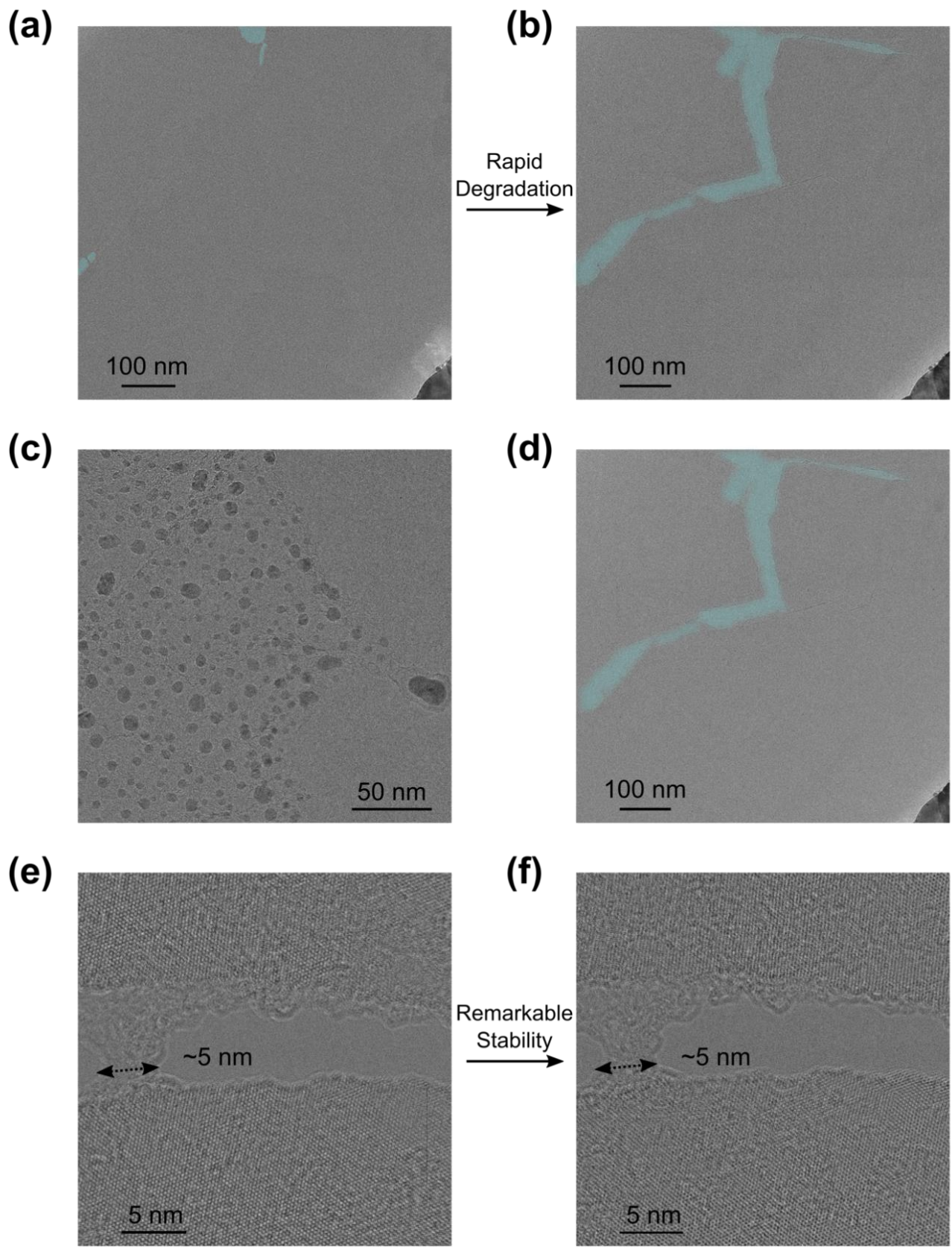

**Figure 4. Void coalescence and stabilization (a)** Low magnification TEM of MoS$_2$ grain boundary region prior to biasing. Voids are colored in blue to aid the eye. **(b)** Same region after an electrical bias is applied. As is apparent, neighboring voids (blue) appear to coalesce to form porous chains. **(c)** After this bias is applied, a number of clusters form within the MoS$_2$ matrix. After a few biasing cycles, applying an electrical bias does not induce a discernible change in the lattice of MoS$_2$. **(d)** Same region in (a) and (b), but the structure largely remains the same after another 5V bias is applied. **(e)** High resolution TEM image of a nanoribbon connecting adjacent to two voids after a few biasing cycles. **(f)** Same region in (e) after a bias of 10V is applied. Once more, no discernible changes in structure are evident and the ribbon retains a lateral dimension of 5 nm.

The differences in activation barriers for the migration between molybdenum atoms and sulfur atoms, as discussed previously, may potentially explain the structural stability of this region.[44,45] Namely, it is likely that with repeated cycles, the line defects mediating this migration process will be continuously shortened through the mechanism suggested by Chen *et al*.[45] In this process, line defects are predicted to slide to voids and molybdenum and sulfur atoms at the edge of these voids migrate into the line defect. This mechanism leads to enlargement of the void and a shortening of the line defect. The shortening of line defects would remove a pathway for the migration of molybdenum atoms and leaves molybdenum mediated vacancy transport as the most favorable route.[50] Because DFT calculations predict the activation energy for molybdenum atoms through this mechanism to be roughly 3 eV, or twice the activation energy barrier for molybdenum atom migration mediated by line defects, the shortening of line defects would significantly slow down the migration of molybdenum atoms. Additionally, though high energy electrons are known to readily create sulfur vacancies, molybdenum atoms tend to be fairly stable and the migration process slows down greatly.[51-53] As such, these resultant structures, such as the narrow ribbons seen in Figure 4e-f, remain stable under electron beam exposure for the small doses used in this study. Our observations suggest that a metastable state is achieved in these samples following several repeated biasing cycles.

**Conclusions**

In this study, we present a method for combining electrical biasing with transmission electron microscopy of monolayer $MoS_2$. We find that voids at the grain boundary of this material play a significant role during electrical biasing. Namely, these neighboring voids coalesce to form long channels likely due to the migration of sulfur and molybdenum atoms. Our results also suggest that the migration of molybdenum atoms serves as the rate-limiting step in this process, which yields a metastable state after prolonged cycling. Through the development of a straightforward methodology for conducting *in situ* electronic measurements on monolayer TMDs and the help of theoretical calculations, we have revealed connections between atomic structure and electronic transport in this system. With this platform, we believe we have created a route for probing a variety of monolayer TMD interfaces in order to better understand and model the complex atomistic interactions at play that give rise to macroscopic properties.

**Methods/Experimental**

*Synthesis of $MoS_2$*

Polycrystalline films of monolayer $MoS_2$ are grown using atmospheric pressure chemical vapor deposition. For this process, 5 mg of $MoO_3$ (Sigma-Aldrich) powder is evenly spread in an alumina boat and covered by two 1cm by 1cm pieces of $SiO_2$/Si wafer (300 nm oxide thickness) that are placed face down. This boat is then placed within a 1-inch diameter quartz tube and into the center of a furnace. Additionally, 130 mg of sulfur pieces (Alfa-Aesar) are placed in another alumina boat that is placed upstream and outside of the furnace. The tube assembly is then purged for 15 minutes with argon gas at a flow rate of 13 sccm. The center of the furnace is then heated to 700°C over a period of 40 minutes and held at 700°C for 3 minutes. During the ramping step, the alumina boat containing the sulfur is moved to a region inside the furnace where the temperature is approximately 150°C. This is done with the help of a magnet and is performed when the temperature in the center region reaches 575°C. Following the growth process, the furnace is allowed to cool naturally.

*TEM device fabrication*

A multi-step transfer method using polycarbonate stamps is used to suspend material over square TEM apertures. First, a gold electrode structure (50 nm thick) is created through standard e-beam lithography on a $SiO_2$ substrate. The patterned substrate is coated with a polycarbonate solution (5% polycarbonate to 95% chloroform by weight) at 2000 RPM for 60 seconds and baked at 120 °C for 1 minute. The substrate is placed in DI water for several seconds until the stamp naturally releases from the substrate, lifting off the gold structure. The polycarbonate stamp supporting the gold structure is then removed from the water and allowed to dry naturally in air for several minutes. Once dry, the stamp is carefully placed on top of two strips of PDMS gel on a glass slide with the gold structure left unsupported and suspended by roughly 1 mm above the glass slide.

Using a micromanipulator, the gold structure is aligned over a TEM window (TEMwindows.com) and brought into contact with the TEM grid. The assembly is gradually heated up to 150°C, which allows the polycarbonate stamp to melt down onto the TEM grid and release from the PDMS support. Once released, the temperature is increased to 170 °C and allowed to bake for 15 minutes. After a cooldown, the structure was put into a chloroform bath for several hours to dissolve the polycarbonate film and then immediately moved to an IPA bath for several minutes to clean off chloroform residue.

This process is repeated for the $MoS_2$, placing the material over the pre-prepared TEM window. Samples are left overnight in chloroform and then annealed in a 95% Ar/5% $H_2$ mixture at 200°C for 2 hours to remove residue prior to TEM imaging. Finally, gold wires are used to connect the electrode pads to the contact points on the TEM mount.

**Associated Content**

A previous version of this paper was uploaded as:

Akshay A. Murthy; Teodor K. Stanev; Roberto dos Reis, Shiqiang Hao, Chris Wolverton, Nathaniel P. Stern, Vinayak P. Dravid. Direct Visualization of Electric Field induced Structural Dynamics in Monolayer Transition Metal Dichalcogenides. 2019, arXiv:1910.02879 [cond-mat.mtrl-sci]. arXiv. https://arxiv.org/abs/1910.02879 (accessed October 8, 2019)

This material is available free of charge *via* the Internet at http://pubs.acs.org.

Supporting_Information.pdf

Experimental details, TEM images of gold nanoparticles, molybdenum clusters, and void formation, DFT calculations, additional strain maps acquired through GPA, and optothermal Raman analysis

SI Video 1.mov

Video demonstrating rupture of nanoribbon connecting two neighboring grains with electrical bias

SI Video 2.mov

Video demonstrating stability of monolayer region with electrical bias following multiple cycles

AUTHOR INFORMATION

**Corresponding author**

*Vinayak P. Dravid: v-dravid@northwestern.edu

**Author Contributions**

The manuscript was written through contributions of all authors. All authors have given approval to the final version of the manuscript. The authors declare no competing financial interest.

**Acknowledgment**

This material is based upon work supported by the National Science Foundation under Grant No. DMR-1507810. This work made use of the EPIC, Keck-II, and SPID facilities of Northwestern University's NU*ANCE* Center, which has received support from the Soft and Hybrid Nanotechnology Experimental (SHyNE) Resource (NSF ECCS-1542205); the MRSEC program (NSF DMR-1720319) at the Materials Research Center; the International Institute for Nanotechnology (IIN); the Keck Foundation; and the State of Illinois, through the IIN. This work also made use of a cryostat platform for confocal microscopy (AttoCube AttoDry 2100, supported by ONR N00014-18-1-2131). A.A.M. gratefully acknowledges support from the Ryan Fellowship and the IIN at Northwestern University. T.K.S. was supported by the Office of Naval Research (N00014-16-1-3055). The authors thank Dr. Anahita Pakzad and Dr. Benjamin Miller from Gatan, Inc, Pleasanton, CA, for the valuable feedback on the usage of K3-IS direct detector.

# Direct Visualization of Electric Field Induced Structural Dynamics in Monolayer Transition Metal Dichalcogenides

*Akshay A. Murthy*[†,§]*, Teodor K. Stanev*[‡]*, Roberto dos Reis*[†,◊]*, Shiqiang Hao*[†]*, Christopher Wolverton*[†]*, Nathaniel P. Stern*[‡]*, Vinayak P. Dravid\**[,†,§,◊]

[†]Department of Materials Science and Engineering, [§]International Institute for Nanotechnology (IIN), [‡]Department of Physics and Astronomy, and [◊]Northwestern University Atomic and Nanoscale Characterization Experimental (NU*ANCE*) Center, Northwestern University, Evanston, Illinois 60208, USA

*Corresponding author

Vinayak P. Dravid:  v-dravid@northwestern.edu

**S1: Experimental details**

*S1.1 Sample Characterization*

S1.1.1 Transmission Electron Microscopy (TEM) and scanning TEM

TEM images were acquired using a JEOL ARM 300F Grand ARM S/TEM operated at 300 kV equipped with Gatan® K3-IS direct electron detector. In particular, HRTEM images were collected in counting mode, which allows us to control the total electron dose applied to the material with high precision, as stacks of 36 images using dose fractionation mode. Typical total dose of 350 e$^-$/A$_2$ was applied. *In situ* videos were collected in counting mode as well, using 75 frames/seconds with frame sizes of 5760 x 4092 pixels. Time compression was applied to reduce data to an easier manageable size keeping significant information of the process. Imaging processing was performed using Gatan Microscopy Suite (GMS) software and custom codes written in MATLAB. Geometrical phase analysis was performed using plug-in written for GMS[1] following methods laid out by Hytch et al.[2]

Scanning transmission electron microscopy (STEM) images and Energy dispersed X-Ray spectroscopy (EDS) were acquired using a Cs-corrected JEOL ARM 200CF operated at 200 kV. High-angle annular dark-field (HAADF) images were obtained with collection angles between 90-200 mrad, with a probe convergent angle of 27mrad and electron current of ~100pA. STEM-EDS data was collected using JEOL Silicon Drift Detector (SDD) with solid angle of 1.7 sr. Simultaneously acquired ADF image were synchronized via Gatan® Digiscan system.

*S1.1.2 Confocal Raman spectroscopy*

Raman spectra were obtained using a Horiba LabRAM HR Evolution Confocal Raman system under laser illumination with a wavelength of 532 nm and a power of 0.5mW.

*S1.1.3 Optothermal Raman thermometry*

Optothermal Raman thermometry was conducted by cooling the fabricated TEM sample using an AttoCube AttoDry 2100 closed-cycle cryostat platform and using the confocal spectroscopy setup to determine the position of the active Raman modes in $MoS_2$ as a function of temperature. A laser wavelength of 532 nm at a power of 0.5 mW was used to illuminate the sample.

*S1.2 Density functional theory calculations*

We used density functional theory (DFT), within the generalized gradient approximation with the PBE functional, periodic boundary conditions and plane wave basis set to obtain the relaxed geometries, as well as the total energies.[3] The PAW pseudopotential have been used for all species. The cutoff energy for the plane-wave basis is 450 eV. For the total energy calculations, all atomic positions are relaxed until the forces exerted on them are less than 0.03 eV/Å. To identify the diffusion minimum energy paths, the climbing nudged elastic band method is used to investigate the barrier for Au diffusion on the surface of monolayer $MoS_2$.[4]

## S2 – Gold nanoparticles

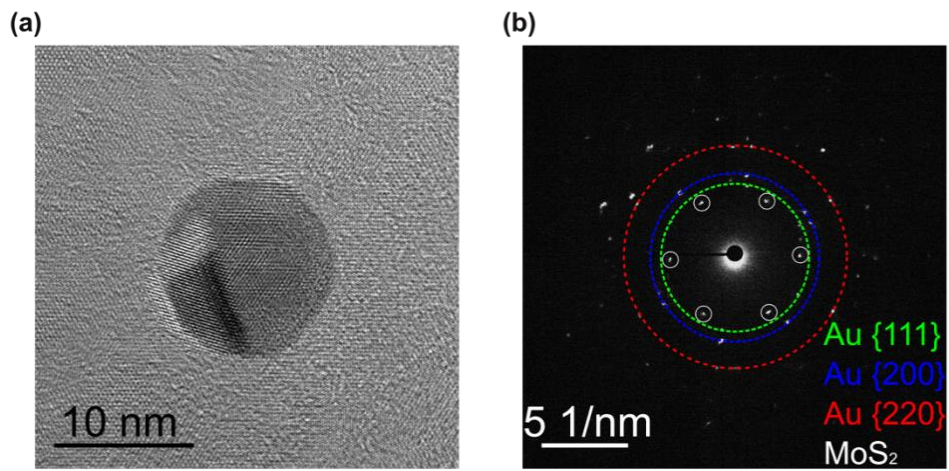

**Figure S2.** (a) High resolution TEM image of a gold nanoparticle in Figure 2b, (b) Selected area diffraction pattern from nanoparticle. The gold diffraction spots lie on Au {111}, Au {200}, and Au {220} diffraction rings, which are indicated by green, blue, and red, respectively. The diffraction spots from MoS$_2$ are circled in white.

## S3: Migration energy of gold atoms on MoS$_2$

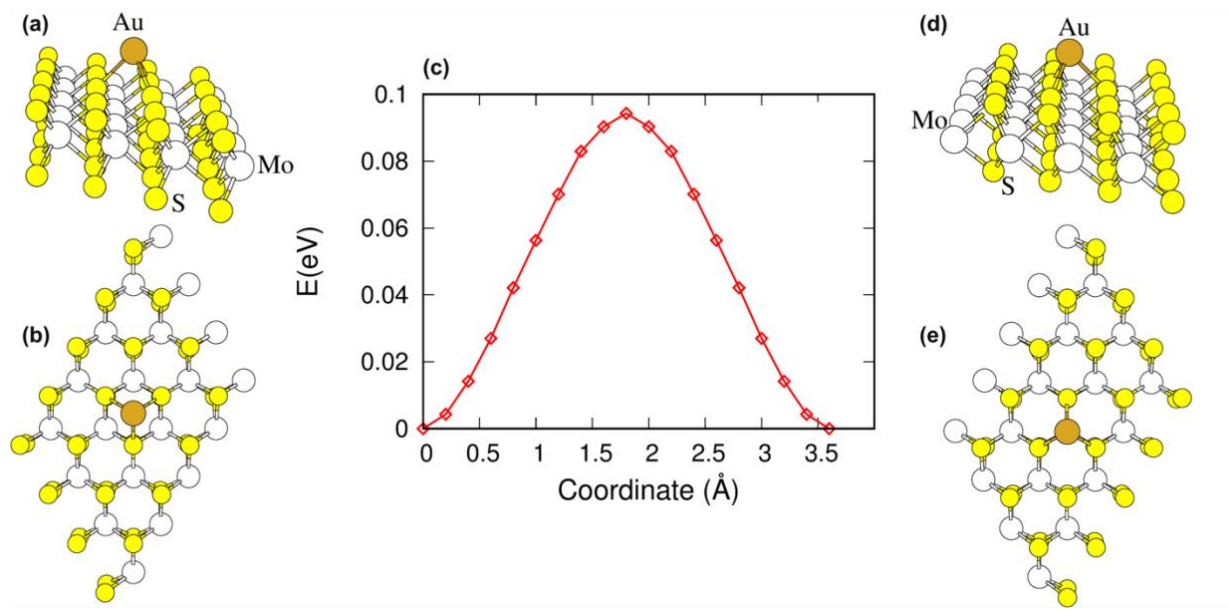

**Figure S3.** (a) Side view of a gold adatom adsorbed on the energetically favorable "hollow" atom position in the MoS$_2$ lattice. (b) Top view of configuration (a). (c) Migration energy barrier profile for diffusion of gold atom from an energetically favorable site to an energetically unfavorable site back to an energetically favorable site. The energy barrier is calculated to be slightly less than 0.1 eV. (d) Side view of a gold adatom adsorbed on the energetically unfavorable Mo atom position in the MoS$_2$ lattice. (e) Top view of configuration (d).

## S4: MoS₂ Grain boundary

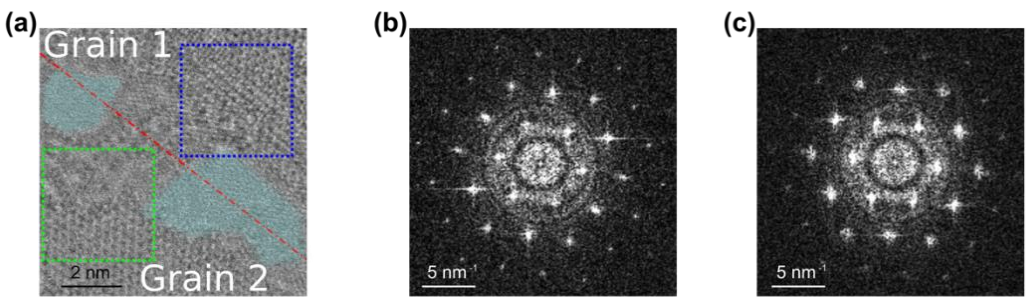

**Figure S4.** (a) High resolution TEM image of MoS$_2$ grain boundary seen in Figure 3a (b) FFT pattern taken from grain 1 – blue box in Figure S4a (c) FFT pattern taken from grain 2 – green box in Figure S4a. The two patterns are offset by 4°.

## S5: Additional strain maps acquired using geometrical phase analysis

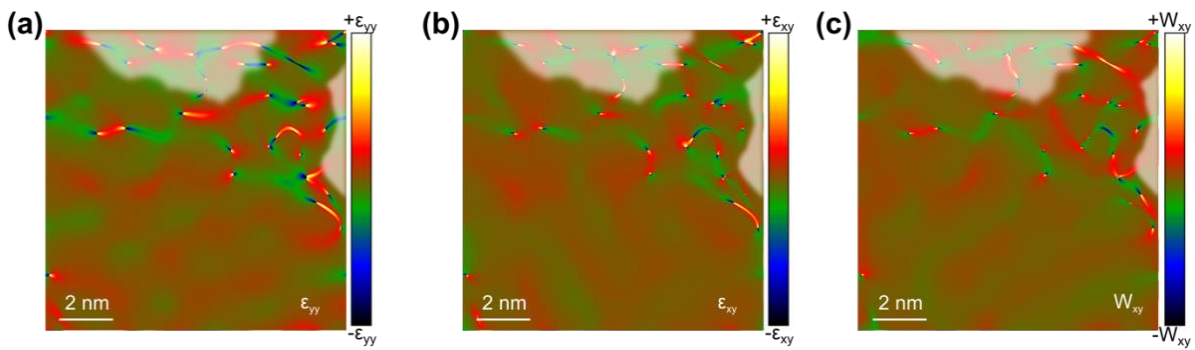

**Figure S5.** Strain maps for the same region seen in Figures 3b-c. (a) Normal strain in y-direction ($\varepsilon_{yy}$) map. (b) Shear strain in x-y plane ($\varepsilon_{xy}$) map. (c) Rotational strain in x-y plane ($W_{xy}$) map.

**S6: Additional images demonstrating impact of biasing**

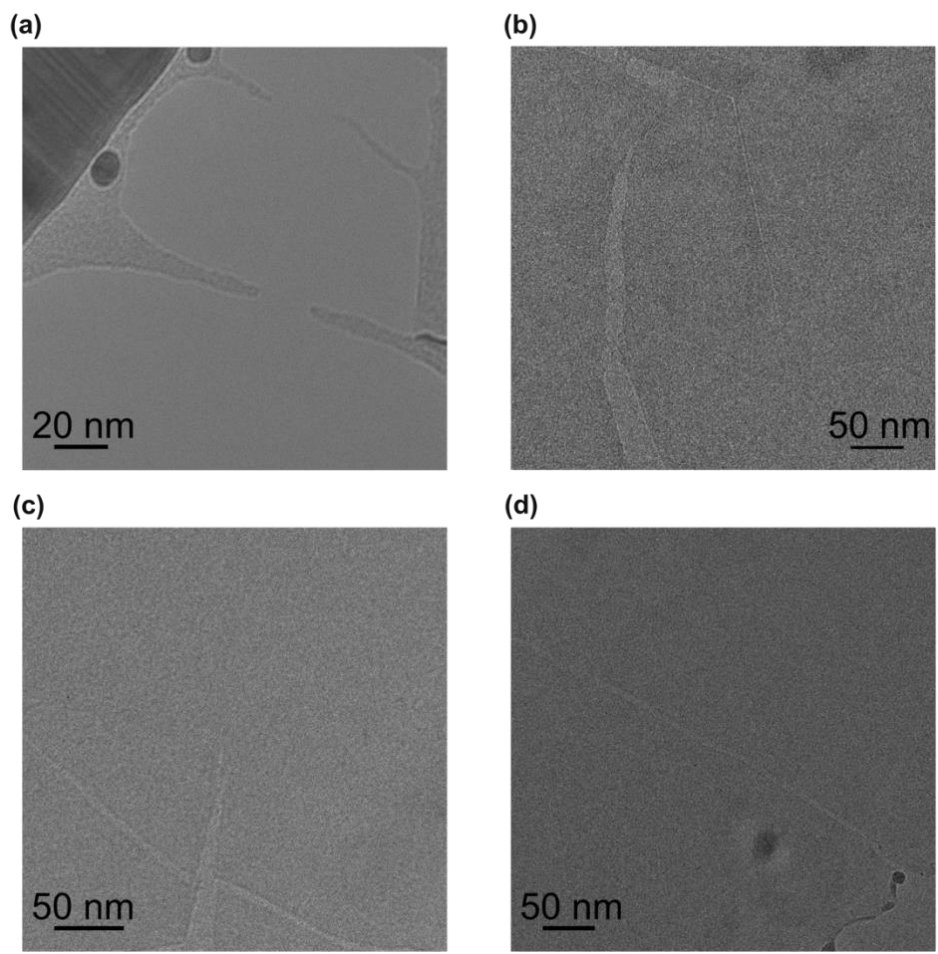

**Figure S6:** Following an applied electrical bias of 5V, broken nanoribbons (a) and coalescence of neighboring voids (b-d) are observed.

## S7: Analysis of molybdenum nanoclusters

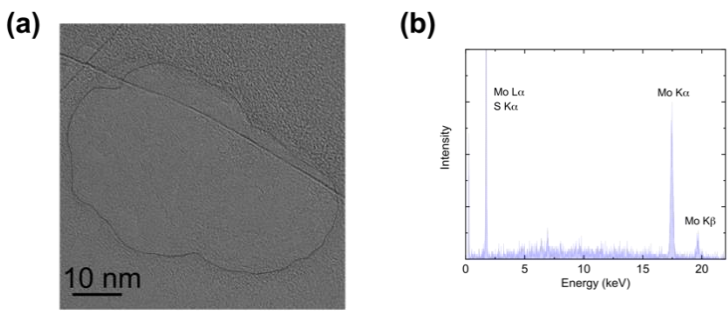

**Figure S7:** (a) High resolution TEM image of molybdenum nanocluster, (b) EDS spectrum of nanocluster.

## S8: Optothermal Raman analysis

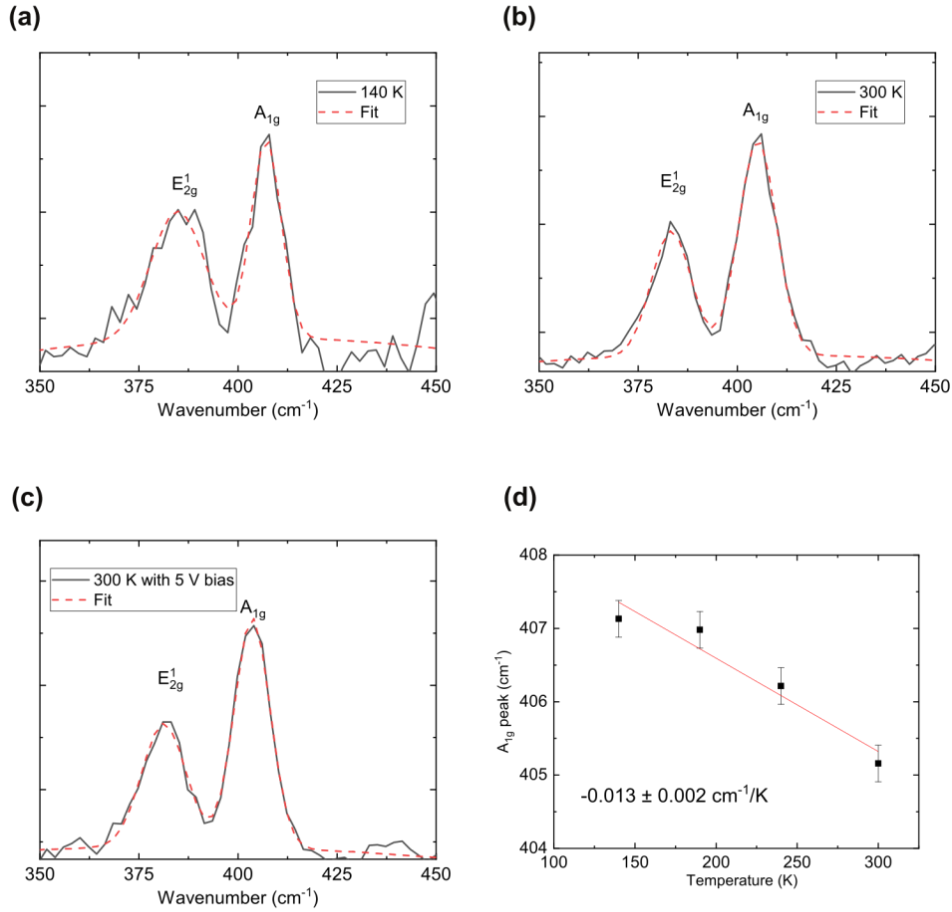

**Figure S8:** (a) Raman spectrum taken from suspended region of $MoS_2$ at 140K with peak fits applied for the $A_{1g}$ and $E^1_{2g}$ modes. (b) Raman spectrum taken from same region at 300K with similar peak fits applied. (c) Raman spectrum taken from the same region at 300K with a 5V applied bias with similar peak fits applied. In this case, the $E^1_{2g}$ mode was centered about 383.7 cm$^{-1}$ and the $A_{1g}$ mode was centered about 403.6 cm$^{-1}$ (d) Location of $A_{1g}$ mode as a function of temperature. The linear regression line has a slope of approximately -0.013 cm$^{-1}$/K with an $R^2$ value of 0.93. Using this relationship and based on a 95% confidence interval, applying a 5V bias is estimated to induce an upper limit in temperature rise of 180K.

## S9: Void coalescence

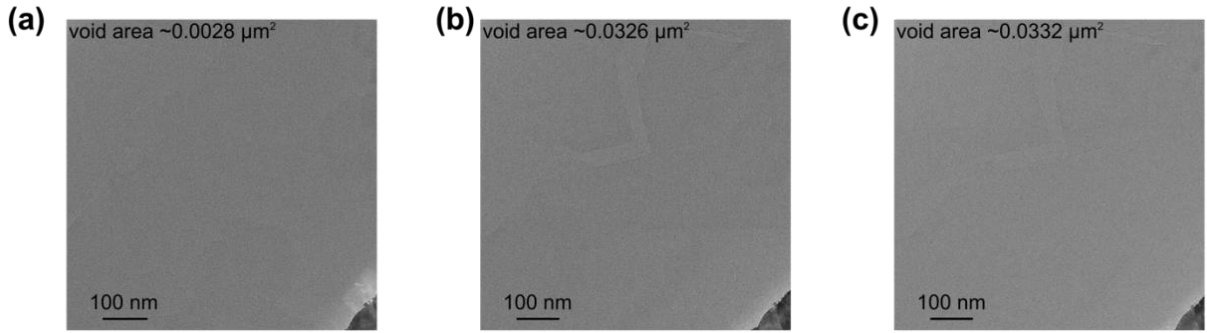

**Figure S9:** (a) Low magnification TEM of MoS$_2$ grain boundary region prior to biasing. (b) Same region after an electrical bias is applied. (c) Same region in (a) and (b), but the structure largely remains the same after another 5V bias is applied. Area of void region dramatically increases after the initial bias, but remains stable following subsequent biasing steps.